\documentclass[12pt]{article}

\usepackage{epsfig}

\begin{document}

\begin{flushright}
hep-ph/0506131  
\end{flushright}

\vskip .4cm
\begin{center}
{\Large \bf Neutrino mass and lepton number violation in the Little Higgs model}
\vskip 1cm
Ashok Goyal \footnote{E--mail :agoyal@iucaa.ernet.in}\\ 
{\em Department of Physics and Astrophysics,University of
Delhi, Delhi-110 007, India}\\
{\em Inter-University Centre for Astronomy and Astrophysics,
Pune-410 007, India}
\end{center}

\begin{abstract}
We study lepton number violation in Little Higgs model and find that the choice of 
putting triplet Higgs vev equal to zero so as not to have any tree level neutrino
 Majorana mass is not natural in the sense that such a term is generated at the one
loop level. We investigate the contribution of exotic lepton number violating terms
on neutrinoless double beta decay, K meson decay and on trilepton production in
 $\nu$-N scattering.  
\end{abstract}

\vskip 1.5cm
\begin{section}{Introduction}
There is compelling evidence for the existence of non zero small
neutrino masses and large neutrino flavour mixing from the Solar,
Atmospheric and Accelerator neutrino data \cite{one}. The SK atmospheric neutrino
and K2K data are best described by dominant
$\nu_{\mu}\leftrightarrow\nu_{\tau}$ vacuum oscillations with best fit
values $|{\Delta M_A}|^2 = 2.1\times10^{-3} eV^2$ and
$sin^2{2\theta_A}=1.0$ at $99.73\%$ CL. The Solar neutrino data
is described by $\nu_e\leftrightarrow\nu_{\mu}$ oscillations with best fit
value ${|{\Delta M_0}}|^2 = 7.9^{+0.6}_{-0.5}\times10^{-5} eV^2$ and
$tan^2{\theta_0}=0.40^{+.09}_{-.07}$. The Troitzk and Mainz tritium
$\beta$-decay experiments \cite{two}  provide information on the absolute
$\bar{\nu_e}$ mass measurement $m_{\bar{\nu_e}}<2.2$ eV at
95\% CL. From the study of anisotropy in the CMBR and large
 scale structure, the WMAP data \cite{three} has severe constraints on the masses
of all active neutrino species $\Sigma m_j<(0.7-1.8) eV$ (95\%
CL). The Katrin experiment expected to start in 2007 is planned to
reach sensitivity $\sim 0.20$ eV (95\% CL) for
$m_{\bar{\nu_e}}$ and the WMAP and future PLANCK experiments and the
data on weak lensing of galaxies by large scale structure may allow
one to determine $\Sigma m_j$ with an uncertainty of (0.04-0.10) eV.

In the minimal Standard Model(SM), the neutrinos are massless because
of the existence of simple Higgs structure which leads to Global
lepton number conservation and forbids the Majorana mass term
$\bar{\nu_L^C}\nu_L$ and the absence of singlet $\nu_R$ forbids Dirac
mass term as well. Thus the masslessness of neutrinos in SM arise due
to restricted particle content of the SM. There are several extensions
of SM that can give non-zero neutrino mass viz. by i) extension of the
lepton sector, ii) extension of the Higgs sector and iii) extension of
both the sectors. The smallness of the neutrino mass in comparison to
the lightest charged lepton mass requires fine tuning of parameters
and is generally considered unnatural. An attractive option currently
favoured is the celebrated See-Saw mechanism which naturally leads to
a set of three light Majorana masses for the three neutrinos along
with the existence of large Majorana mass right-handed electro-weak
 singlet neutrinos. It is thus imperative to go beyond the SM and look
for new physics. Neutrino mass generation is not the only problem
afflicting SM. The other problem is the so called Hierarchy problem,
that is enormous difference between the electro-weak and GUT/Planck
scale. The precision electro-weak data prefers the existence of light
Higgs and thus SM with light Higgs can be considered as an effective
theory valid to a high scale perhaps all the way to GUT/Planck scale
whereas the Higgs mass is not protected and gets quadratically
divergent contribution to its mass and requires fine tuning. One way
out of this difficulty is to invoke Supersymmetry where quadratic
divergences in Higgs mass are cancelled by the corresponding fermionic
contribution.In fact there are several post SM scenarios devised to ameleorate
 this difficulty namely : SUSY, Left-Right symmetric gauge theories, GUTS,
 theories of extra dimensions etc. All these theories can naturally 
incorporate Majorana neutrinos.

 Recently there has been a proposal to consider Higgs
fields as Nambu-Goldstone bosons of a Global symmetry which is
spontaneously broken at some high scale f by acquiring vacuum
expectation value (vev). The Higgs field gets a mass through
electro-weak symmetry breaking triggered by radiative corrections
leading to Coleman-Weinberg type of potential. Since the Higgs is
protected by approximate Global symmetry, it remains light and the
quadratic divergent contributions to its mass are cancelled between
particles of the same statistics. The Littlest-Higgs (LH) model \cite{four} is a
minimal model of this genre which accomplishes this task of cancelling
quadratic divergence to one loop order with a minimal matter
content. The LH model consists of an SU(5) non-linear sigma model
which is broken down to SO(5) by a vacuum expectation value f. The
gauged subgroup $[SU(2)\times SU(1)]^2$ is broken at the same time to
diagonal electro-weak SM subgroup$ [SU(2)\times SU(1)]$. The new heavy states in
this model consist of vector 'top quark' which cancels the quadratic
divergence coming from the SM top quark along with the new heavy gauge
bosons $(W_H,Z_H,A_H)$ and a triplet Higgs $\Phi$, all of masses of
order f and in the TeV range. The effect of these new states on
electro-weak precision parameters has been studied to put constraints
on the parameters of the model \cite{five}. In particular no stringent limit on$ v'$
the vev of triplet Higgs exist apart from the bound
$\frac{v'^2}{v^2}<\frac{v^2}{16f^2}$ from demanding positive definite
mass for the trplet Higgs. It should be noted that $v'$ can be made to
vanish by tuning the parameters so that the triplet Higgs does not
couple to standard doublet Higgs.

The existence of triplet Higgs allows the lepton number violating
interaction  
\begin{equation}
 L_{LV}=-g_{\Phi ll}{\bar l}_L^C \tau_2 \vec{\tau}.\vec{\Phi}
l_L +hc
\label{eq:1}
\end{equation}
which is invariant under the electro-weak gauge symmetry of the LH model.
 After electro-weak symmetry breaking, such an interaction will
 generate Majorana masses for the left handed neutrino of order
 $g_{\Phi ll}v'$ and will allow for lepton number violating processes
 like neutrinoless double $\beta$-decay. In order to achieve neutrino
 Majorana mass consistent with current bounds on neutrino masses from
 neutrino oscillation data which does not differentiate between the
 Majorana and Dirac masses and from costraints from neutrinoless
 double $\beta$-decay experiments, either the coupling $g_{\Phi ll}$
 or the triplet Higgs vev $v'$ should be unnaturally small ie
.$g_{\Phi\nu\nu}.v' \sim 10^{-10}$ GeV. The vacuum
 expectation value $v'$ can be taken to be zero as discussed above, on
 the other hand putting the coupling $g_{\Phi ll}$ equal to zero by
 hand ammounts to an adhoc imposition of lepton number
 conservation. Vanishing of $v'$ admittedly requires fine tuning but is
 perhaps a small price to be paid to achieve non-zero neutrino mass
 and lepton number violation in the attractive frame work of LH model
 without any further extension. 
\end{section}

\begin{figure}[htb]
\begin{center}
\epsfig{file=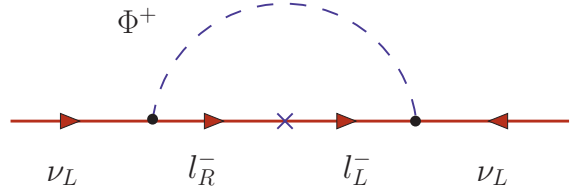,width=.45\textwidth}
\caption{Neutrino Majorana mass at one loop level}
\end{center}
\label{fig:1}
\end{figure}

\begin{figure}[htb]
\begin{center}
\epsfig{file=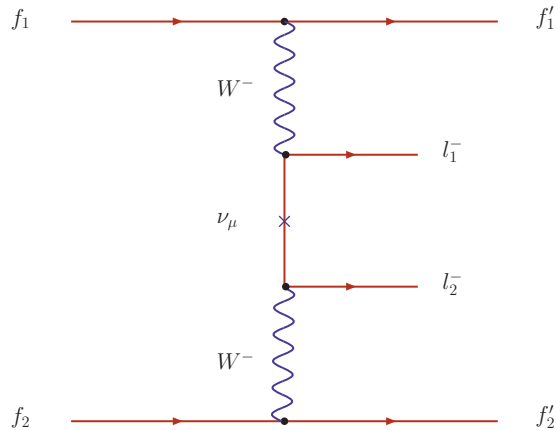,width=.45\textwidth}
\caption{Generic diagram for $\triangle L = 2$ Majorana neutrino mediated processes} 
\end{center}
\label{fig:2}
\end{figure}

\begin{figure}[htb]
\begin{center}
\epsfig{file=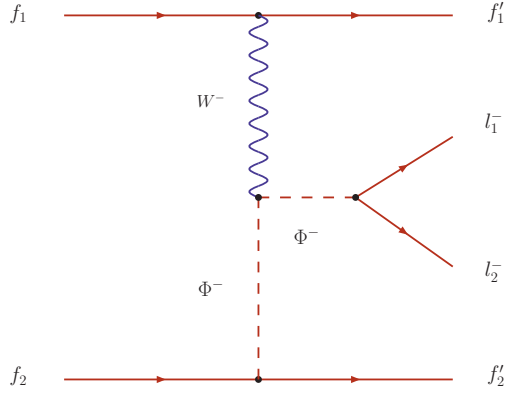,width=.45\textwidth}
\caption{Generic diagram for $\triangle L = 2$ processes mediated by
  doubly charged triplet Higgs  } 
\end{center}
\label{fig:3}
\end{figure}

\begin{section}{Neutrino mass and lepton number violating processes}
In the scenario in which the triplet Higgs does not develop a vev, the
neutrinos will get masses through loop diagrams shown in Fig.1 in
which one of the the black circles stand for the lepton number violating vertex
[1] and the cross represents charged lepton mass insertion. The
diagram is log divergent. The divergence can be absorbed in tree level mass
 term arising from nonzero vev $v'$ which in turn would mean that the choice
 $v'$ is not natural. In fact the above diagrams will generates nonzero
 vev through  $g_{\Phi\nu\nu}\times\bar\nu_L\nu_L$ term in (1).
Some of the important lepton number violating processes that we consider are:

\begin{itemize}
\item[1] The neutrinoless double $\beta$-decay process
  $(N,Z)\longrightarrow (N-2,Z+2) +e^-+e^-$
\item[2] K meson decay $K^+\longrightarrow \pi^- \mu^+ \mu^+$
\item[3] $\tau$ decay  $ \tau^-\longrightarrow e^+ +\pi^- +\pi^-$ 
\item[4] Nuclear muon-positron conversion $^{32}S +
  \mu^-\longrightarrow ^{32}S +e^-$   

\item[5] Trimuon production in neutrino factories and in the ultra
 high energy cosmic  ray neutrino interaction processes $\nu
 N\longrightarrow \mu^-\mu^+\mu^+ X$ 
\end{itemize}

The above lepton number violating $\Delta$L=2 processes mediated by 
Majorana neutrinos are depicted in fig.2 and the matrix elements are
proportional to neutrino flavor mixing matrix elements and lepton
number violating Majorana neutrino mass insertion. In the presence of
lepton number violating interaction in the LH-model, the dominant
generic
diagram is through the decay of doubly charged Higgs into lepton pair
and is shown in fig.3. For the neutrinoless double $\beta$ decay these
diagrams are characterised by the following couplings 
$$M_{\nu} \sim (\frac{g}{2\sqrt2})^4 \frac{1}{M_W^4}
\frac{<m>_{l_1l_2}}
{<q^2>}$$
$$M_{\Phi} \sim (\frac{g}{2\sqrt2})^2 \frac{1}{M_W^2}\frac
{8}{M_{\Phi}^4}m_{f_2}(\frac{v}{f}-2s_+)g_{\Phi l_1l_2}$$

In the case of $K^+ \longrightarrow \pi^-\mu^+\mu^+$ decay, the
dominant s-channel diagrams yield the amplitude

\begin{equation}
M_{\nu}=2G_F^2f_Kf_{\pi}(V_{ud}V_{us})^*\frac{M_{\nu}}{q^2-m_{\nu}^2}
p_{K\mu}p_{\pi\nu}[L^{\mu\nu}(k_1,k_2)-L^{\mu\nu}(k_2,k_1)]
\end{equation}
where $L^{\mu\nu}(k_1,k_2)=\bar v(k_1)\gamma^{\mu}\gamma^{\nu}P_Rv^C(k_2)$
and
\begin{equation}
M_{\Phi}=\sqrt2G_Ff_Kf_{\pi}(V_{ud}V_{us})^*\frac{M_K^2}{M_{\Phi}^4}\frac{1}{f}
p_K.p_{\pi}[L(k_1,k_2)-L(k_2,k_1)]  
\end{equation}
where $L(k_1,k_2)=\bar v(k_1)P_Rv^C(k_2)$
Decay rates can be calculated and we get 
\begin{equation}
\Gamma_{\nu}= G_F^4f_K^2f_{\pi}^2
m_K^5|V_{ud}V_{us}|^2\frac{<m>_{\mu}^2}{M_K^2} I_{\nu}
\end{equation}
\begin{equation}
\Gamma_{\Phi}=G_F^2f_K^2f_{\pi}^2 \frac{
m_K^2}{f^2}\frac{m_K^8}{m_{\Phi}^8} m_K|V_{ud}V_{us}|^2 g_{\Phi\mu\mu}^2 I_{\Phi}
\end{equation}
where $I_{\nu}$ and $I_{\Phi}$ are dimensionless phase space
integrals.
\\

In the trimuon production process $ \nu_\mu N \longrightarrow 
\mu_-\mu_+\mu_+X$, the relative effective couplings in the amplitude
from the two diagrams are 
               
$M_\nu \sim G_F^2 \frac{m_\mu}{<q^2>}$ 
and
 $M_{\Phi}\sim G_F(\frac{m_q m_\mu}{M_{\Phi}^4})
\frac{1}{v}(\frac{v}{f} -2s_+ )g_{\Phi \mu \mu}$ respectively.
\end{section}
\begin{section}{Summary and Conclusions}
The above lepton number violating processes mediated by Majorana
neutrino exchange have been studied in the litrature \cite{six} and limits 
on the Majorana neutrino mass placed. The best limits on Majorana neutron
mass come from neutrinoless double $\beta$-decay of $^{76}Ge$ which
gives $<m>_{ee} < 0.3$ eV. Limits on $<m>_{e \mu}$ varies from 17 MeV to
$<$ 1.3 TeV from muon-positron conversion in sulphur. Limits on
$<m>_{\mu\mu}$ and $<m>_{\tau l}$ are very weak and are obtained from
K and $\tau$ decays and are $ <$ 0.48 and $<$ 12 TeV respectively.In the
presence of lepton number violating interactions in the LS model, the
corresponding decay rates and branching ratios need not be supressed
by the low neutrino masses and can indeed be large. They are
proportional to the coupling $g_{\Phi ll}$ and the most stringent
limit on the coupling comes from the neutrino-less double
$\beta$-decay.
It should be emphasised that the experimental bounds on the lepton
number violating processes discussed here except for neutrinoless
double $\beta$-decay are so weak that any meaningful bounds on either
the light majorana neutrino masses or couplings $g_{\Phi ll}$ do not
exist. Alternatively, if lepton number violating processes are indeed
observed, they would strongly point towards the existence of exotic
lepton number violating interactions such as discussed above arising
in the LS model. 

\end{section}

\end{document}